\documentclass[prl,twocolumn,superscriptaddress,floatfix]{revtex4-1}

\usepackage{graphicx}
\usepackage{multirow}

\begin{document}

\title{Ferroelectric PbTiO$_{3}$/SrRuO$_{3}$ superlattices with broken inversion symmetry}
\author{S.J. Callori}
\author{J. Gabel}
\affiliation{Department of Physics and Astronomy, Stony Brook University, Stony Brook, NY 11794-3800 USA}
\author{Dong Su}
\affiliation{Center for Functional Nanomaterials, Brookhaven National Laboratory, Upton, NY, USA}
\author{J. Sinsheimer}
\author{M.V. Fernandez-Serra}
\affiliation{Department of Physics and Astronomy, Stony Brook University, Stony Brook, NY 11794-3800 USA}
\author{M. Dawber}
\email{matthew.dawber@stonybrook.edu}
\affiliation{Department of Physics and Astronomy, Stony Brook University, Stony Brook, NY 11794-3800 USA}

\begin{abstract}

We have fabricated PbTiO$_{3}$/SrRuO$_{3}$ superlattices with ultra-thin SrRuO$_{3}$ layers. Due to the superlattice geometry, the samples show a large anisotropy
in their electrical resistivity, which can be controlled by changing the thickness of the PbTiO$_{3}$ layers. Therefore, along the ferroelectric direction, SrRuO$_{3}$ layers can act as dielectric, rather than metallic, elements. We show that, by reducing the concentration of PbTiO$_{3}$, an increasingly important effect of polarization asymmetry due to compositional inversion symmetry breaking occurs. The results are significant as they represent  a new class of ferroelectric superlattices, with a rich and complex phase diagram.
 By expanding our set of materials we are able to introduce new behaviors that can only occur when one of the materials is not a perovskite titanate. Here, compositional inversion symmetry breaking in bi-color superlattices, due to the combined variation of A and B site ions within the superlattice, is demonstrated using a combination of experimental measurements and first principles density functional theory.

\end{abstract}

\maketitle

Artificially layered perovskite oxide superlattices provide many opportunities to develop systems with novel and tunable properties\cite{Dawber08,Zubko11}.
As far as ferroelectric superlattices are concerned, the insulating titanium perovskite oxides (e.g. PbTiO$_{3}$, BaTiO$_{3}$, CaTiO$_{3}$ and SrTiO$_{3}$) have to date been the most popular ``building blocks'',  but the need for new functionalities, particularly related to magnetism,  requires the use of a wider set of materials and a deep understanding of the new physical phenomenon related to interfaces. 
In this letter an unconventional approach is demonstrated: we use a material that is normally metallic to play the role of a dielectric, in a ferroelectric-dielectric superlattice.

The much-studied compound SrRuO$_{3}$ provides the proof of concept that metallic magnetic oxides can transform into thin-film dielectric components in certain heterostructures.
 In bulk, SrRuO$_{3}$ has the distorted perovskite orthorhombic Pnma structure, is metallic and is ferromagnetic below Tc = 160 K \cite{Allen96,Singh96,Mazin97,Dodge00,Capogna02}.
It is also a commonly used electrode material for oxides, and the interface with ferroelectric oxides has been much studied \cite{Junquera03,Gerra06,Stengel06,AguadoPuente08,Shin10}. 
However, SrRuO$_{3}$ becomes insulating in layers of thickness less than 4 unit cells; this behavior has been observed in thin films \cite{Toyota05,Schultz09,Xia09} and in SrTiO$_{3}$/SrRuO$_{3}$ superlattices\cite{Izumi98,Kumigashira08}. 
First-principles investigations \cite{Zayak06,Rondinelli08,Mahadevan10} and experiment \cite{Herranz03,Kim05,Moore07,Grutter10,Ziese10,Choi10} indicate that epitaxial strain, size effects, chemical pressure, surface reconstruction and interaction with the substrate may all play an important role in the observed behavior.
In a recent contribution Verissimo-Alves et al. \cite{VerissimoAlves12} showed from first principles calculations that a highly confined 2DEG is formed at the interface in SrTiO$_{3}$/SrRuO$_{3}$ superlattices. 
We will show in this letter that a similar effect occurs in PbTiO$_{3}$/SrRuO$_{3}$ superlattices, but that in the direction perpendicular to the interfaces PbTiO$_{3}$/SrRuO$_{3}$ superlattices containing  single unit cell  layers of SrRuO$_{3}$ are insulating and can be ferroelectric.

A second motivation for creating PbTiO$_{3}$/SrRuO$_{3}$  superlattices is that, as they have both A and B site variation, inversion symmetry can be compositionally broken \cite{Sai00,Sai01}. 
As a result, an asymmetry is introduced in the ferroelectric double-well potential which can lead to ``self-poling" materials. 
Self-poling ferroelectric materials are useful in piezoelectric applications where the desired mode of operation is to apply an electric field either with or against a fixed polarization direction to achieve, respectively, an expansion or contraction of the material.
In a tri-color superlattice compositional breaking of inversion symmetry can occur with only A or B site variation. The effect has been seen with A site variation in tri-layer superlattices containing BaTiO$_{3}$, CaTiO$_{3}$ and SrTiO$_{3}$ \cite{Warusawithana03,Lee05},  and shown to be an appealing route towards magnetoelectric materials  with tri-color variation on the B site\cite{Hatt07,Yamada08}, but until now this behavior has not been shown experimentally in bi-color  superlattices.

Epitaxial growth of PbTiO$_{3}$/SrRuO$_{3}$ superlattices can be achieved on SrTiO$_3$ substrates as both PbTiO$_{3}$ and SrRuO$_{3}$ have in-plane lattice parameters close to that of SrTiO$_3$, which at room temperature is cubic with a=3.905 \AA. 
At room temperature bulk PbTiO$_{3}$ is tetragonal (a=3.904 \AA, c=4.15 \AA) and  orthorhombic SrRuO$_{3}$ can be considered as pseudo-cubic with a=3.93 \AA.
For this study, the n/1 PbTiO$_{3}$/SrRuO$_{3}$ (n unit cells (u.c.)  PbTiO$_{3}$/1 u.c. SrRuO$_{3}$) superlattices were deposited using off-axis RF magnetron sputtering on (001) SrTiO$_3$ substrates, which had been treated with buffered HF and annealed to ensure TiO$_{2}$ termination.
The SrRuO$_{3}$ layer thickness was grown to 1 u.c. for all the samples considered here, with the aim being that the SrRuO$_{3}$ layers should act as dielectrics, rather than metals.
By contrast the thickness of the PbTiO$_{3}$ layer was changed from sample to sample, so that the relative effect of bulk ferroelectricity vs.  interfacially driven compositional inversion symmetry breaking could be assessed.
The total number of bilayers in the superlattice was varied from one sample to another so that the total thickness of each sample was in each case as close as possible to 100nm.
Growth rates for the two materials within the superlattice were obtained from x-ray diffraction measurements performed on many preliminary samples. 
Bottom SrRuO$_{3}$ electrodes (20 nm in thickness) were deposited \emph{in situ} for the samples used for electrical measurements  and gold top electrodes were added to the samples post-deposition. 
The superlattices were grown at a temperature of 550$^{o}$ C and the SrRuO$_{3}$ electrodes were grown at 620$^{o}$ C.

Experimentally, it has been demonstrated by Rjinders et al \cite{Rijnders04} that when grown by pulsed laser deposition the termination of a SrRuO$_{3}$ film is affected by both deposition conditions and layer thickness and that as a film grows on Ti terminated SrTiO$_{3}$ there is a conversion from a RuO$_{2}$ to a SrO termination layer.  
At a growth temperature of 700$^{o}$C  this transition already occurs for a single unit cell SrRuO$_{3}$ layer, but at lower deposition temperatures, this transition occurs later in the growth, as the RuO$_{2}$ layer is comparably more stable at these conditions.
As a consequence of the small thickness of our SrRuO$_{3}$ layers, the low deposition temperatures used in our process, and the different kinetic regime of sputtering as compared to pulsed laser deposition, a RuO$_{2}$  termination of our SrRuO$_{3}$ layers, may still be possible.

\begin{figure}[h]
  \includegraphics[width=8.5cm]{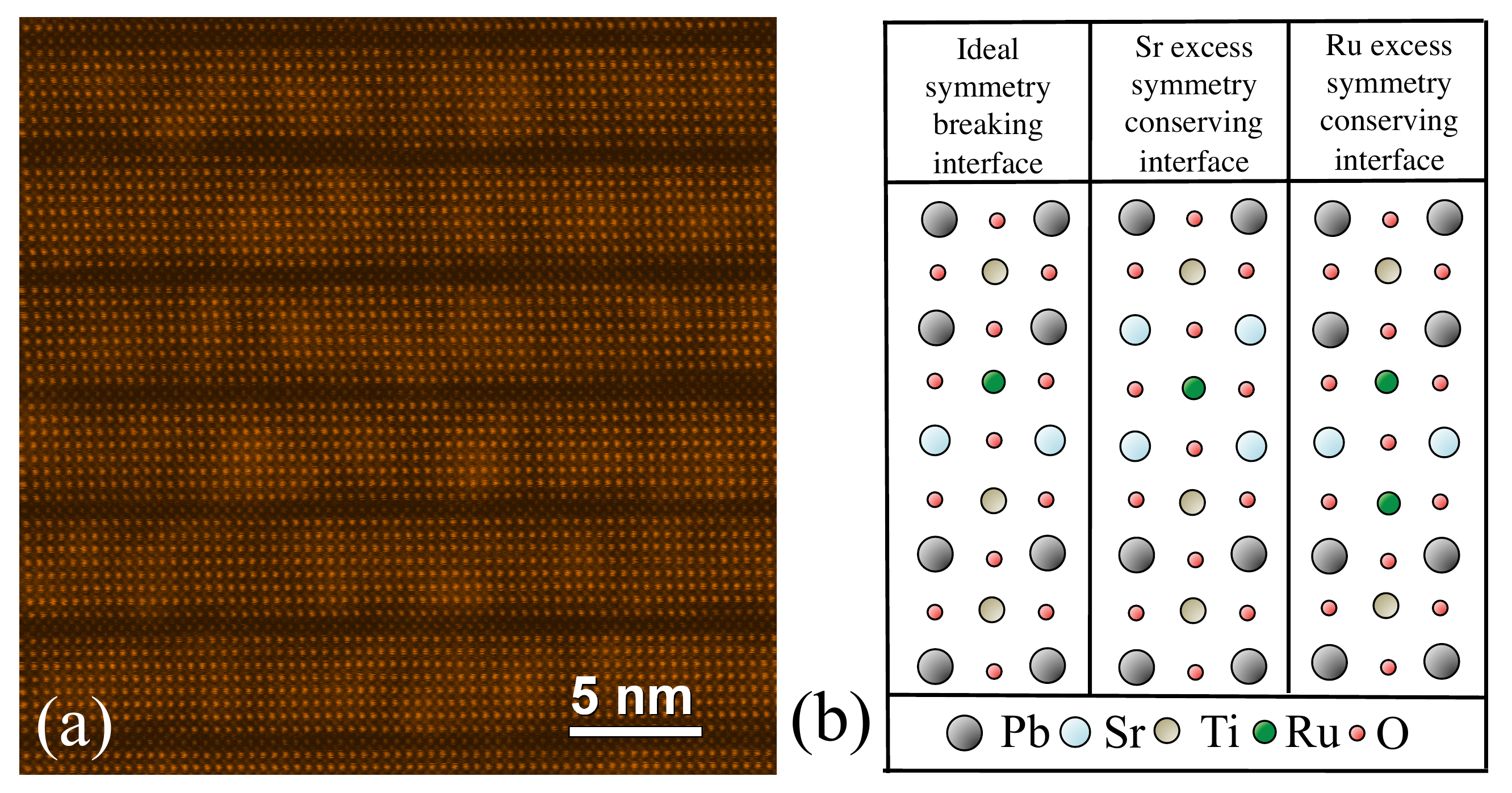} 
  \caption{\textit{(a) A HR-STEM image  of an  8/1 PbTiO$_3$/SrRuO$_3$ superlattice. (b) The three types of interfaces considered theoretically in this paper,  illustrated for the case of a 3/1 PbTiO$_3$/SrRuO$_3$ superlattice.}}\label{TEM}
\end{figure}

The epitaxial growth  of our samples was confirmed by x-ray diffraction and high resolution scanning transmission electron microscopy (HR STEM).
Fig. \ref{TEM}(b) shows a HR STEM cross section of an 8/1 PbTiO$_{3}$/SrRuO$_{3}$ superlattice.
The PbTiO$_{3}$ layers are the brightest because of the high atomic number of Pb, while the SrRuO$_{3}$ layers are less bright than   PbTiO$_{3}$,  but have enhanced brightness compared to SrTiO$_{3}$ (not shown) because the Ru ion has a higher atomic number than Ti.
In addition to the STEM image shown, we also carried out STEM-EELS line scans \cite{SI}, which support the ideality of our grown structures.

Although the interface most likely to form in PbTiO$_{3}$/SrRuO$_{3}$ superlattices if the atoms in the material are deposited in the same ratio as the parent targets breaks inversion symmetry it is possible to consider interfaces that might form which do not. 
In Fig. \ref{TEM} (b) we show three kinds of interfaces which we have studied using first principles calculations, which we illustrate for the case of  a 3/1 superlattice.
The ideal structure which breaks inversion symmetry is the first from the left in Fig. \ref{TEM} (b).  
In order to study the significance of the symmetry breaking effect, we also simulated two different kinds of interface which conserve symmetry, one in which one Pb-O plane has been replaced by a Sr-O plane,  which is shown in the center of Fig. \ref{TEM} (b),  and another, (less likely due to the high volatility of Ru), unit cell in which a Ti-O  plane has been replaced by a Ru-O plane is shown on the right.

\begin{figure}[h]
 \includegraphics[width=8.5cm]{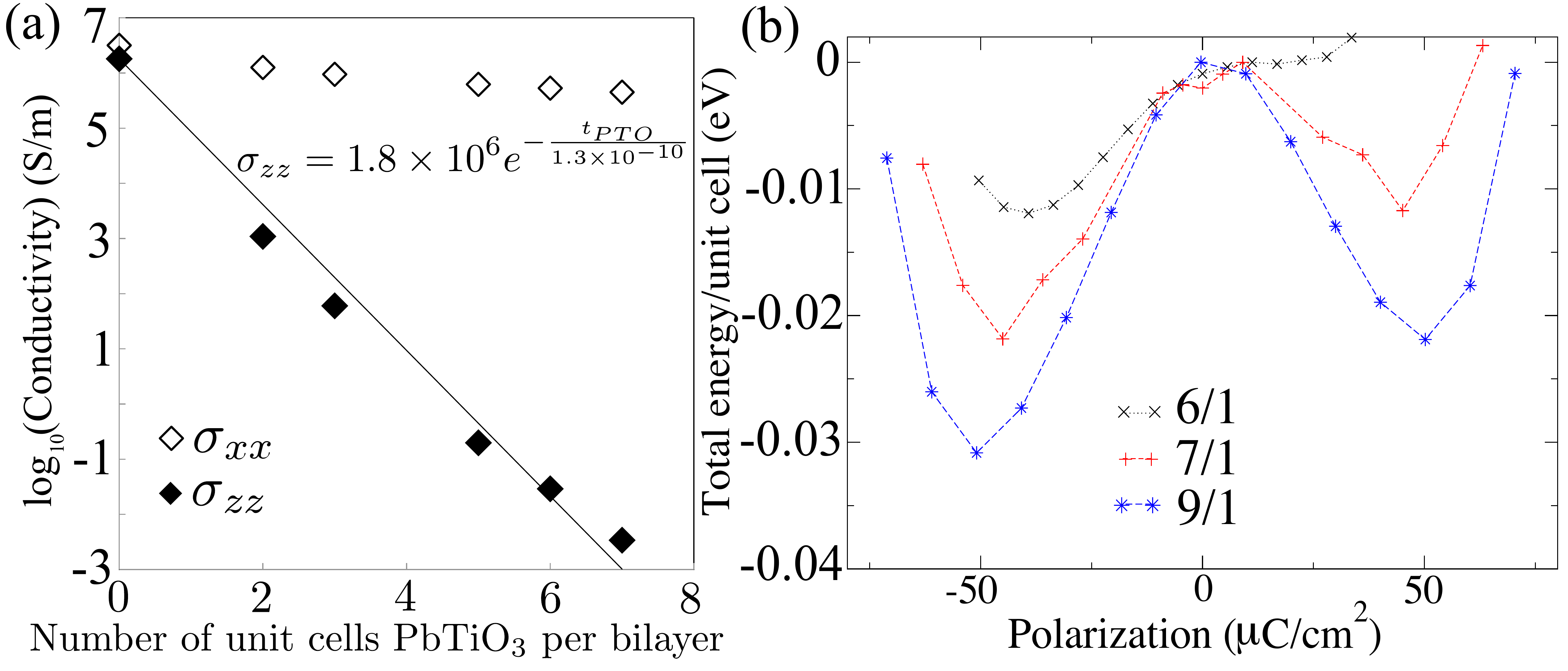}
 \caption{\textit{(a) Calculated conductivity both in-plane and out of plane plotted a function of the number of unit cells of PbTiO$_{3}$ included in each bilayer (with zero corresponding with SrRuO$_{3}$ strained in-plane to the SrTiO$_{3}$ lattice parameter). (b) Calculated total energy per unit cell as a function of polarization.}}\label{DoubleWell}
\end{figure}

We investigated the 3 kinds of interface shown above using first principles calculations. 
These were performed using density functional theory, using  a basis of numerical atomic orbitals as implemented in the {\sc siesta} code. We used the same basis set and pseudopotentials as Verissimo-Alves et al. \cite{VerissimoAlves12}.
We studied the influence of spin polarization, the use of the generalized gradients approximation within the commonly used Wu-Cohen parametrization\cite{Wu06}, and the effect of correlations within the LDA+U and LSDA+U approximations\cite{Sanvito04}.
The different approximations used can affect the electronic properties of the metallic layer along the parallel direction. 
However the electrical anisotropy is mostly dependent on the superlattice periodicity, and therefore results are computed using the local density approximation (LDA), which is the best method to characterize both the structural and electronic properties of the superlattices.
Full details regarding the calculations can be found in the supplemental materials to this letter\cite{SI}.
We examined the electrical conductivity of the superlattices both in-plane ( $\sigma_{xx}$) and out-of-plane ($\sigma_{zz}$) by calculating the diagonal elements of the conductivity tensor within the relaxation time approximation to the Boltzman transport equation\cite{Allen88}. The anisotropy in $\sigma$ is fully determined by the anisotropy of the Fermi surface geometry, as determined by  $\sigma_{\alpha\alpha}=-e^{2}\tau\sum_{k}v_{k\alpha}^{2}\delta\left(\epsilon_{F}-\epsilon_{k}\right)$.
The relaxation time, $\tau$ is the only variational parameter in the expression, we choose $\tau=1.3\times10^{-14}$ s after experimental results in bulk SrRuO$_{3}$\cite{Santi97}.
This approximation ignores the anisotropy of the electron-phonon scattering, although this is known to be a much smaller effect\cite{Allen88}.
The results shown in Fig. \ref{DoubleWell} (a) are for the ideal interfaces, but these quantities were also calculated for the two other cases and are qualitatively similar, again an indication that the anisotropy is almost fully determined by the inter- SrRuO$_{3}$ layer distance.
We find, in agreement with ref. \cite{VerissimoAlves12}, that the electrons in the single unit cell layers of SrRuO$_{3}$ are confined to that layer, so while the in-plane conductivity, $\sigma_{xx}$, of the structures does not change dramatically as the spacing between the layers is varied the out-of-plane conductivity, $\sigma_{zz}$,  decreases exponentially with a  characteristic length of 1.3$\mathrm{\AA}$ as the thickness of the PbTiO$_{3}$ layers is increased. 

In the two-component PbTiO$_{3}$/SrRuO$_{3}$ superlattices in which the inversion symmetry is broken by the ideal interface structure, our calculations predict a self-poling behavior.
In Fig. \ref{DoubleWell} (b) we show the energy of the superlattice as a function of the polarization.
We computed the total energy of the system for atomic
displacements along the line $\vec{r}=\vec{r}_{P_+} +u(\vec{r}_{P_-} -\vec{r}_{P_+})$, interpolating linearly between the two minima of the energy.    
These two minima are characterized by two different polarization states, $P_{up}$ (higher energy minimum) and $P_{down}$ (lower energy minimum).
As shown in Table \ref{secondtable}, the simulations for the ideal interface show that, as the PbTiO$_{3}$ layer thickness is reduced, there is an increasingly large difference in the values of the stable up and down polarizations, until for the 5/1 superlattice, when the potential well has just one minimum, only the down polarization is stable.

\begin{table}[!h]
\begin{tabular}{|c|c|c|c|c|}
  \hline
  Superlattice & \multicolumn{4}{c|}{Stable Polarization ($\mathrm{\mu C/cm^{2}}$)} \\ \hline
  & \multicolumn{2}{c|}{Ideal} & Sr Excess & Ru excess \\ \hline
                            & $P_ {down}$ & $P_{up}$ & $P$ & $P$ \\ \hline

  (PbTiO$_{3}$)$_{5}$(SrRuO$_{3}$)$_{1}$ & 35.8 & unstable & 2.0 & 23.3 \\
  (PbTiO$_{3}$)$_{6}$(SrRuO$_{3}$)$_{1}$ & 39.2 & 16.8 & 16.4 &  32.5\\
  (PbTiO$_{3}$)$_{7}$(SrRuO$_{3}$)$_{1}$ & 45.0 & 45.1 & 27.2 &  44.2 \\
  (PbTiO$_{3}$)$_{9}$(SrRuO$_{3}$)$_{1}$ & 50.9 & 50.2 & 42.9 & 52.7 \\
  \hline

\end{tabular}
\caption{\textit{Calculated stable polarization magnitudes (DFT LDA) in the up and down directions for  a selection of ideal superlattices.}}
\label{secondtable}
\end{table}

The preference of one polarization state over another for superlattices with broken compositional inversion symmetry was seen in all of the calculation schemes used. 
In the spin polarized calculations for samples with single unit cell layers of SrRuO$_{3}$ the spin polarization is not affected by the direction of the polarization.
However, in simulations of superlattices with symmetry breaking interfaces that have SrRuO$_{3}$ layers thicker than one unit cell the magnetization is different for the two polarization directions.
 Although it is not the focus of the present paper, this  finding demonstrates the potential for the compositional breaking of inversion symmetry at the PbTiO$_{3}$/SrRuO$_{3}$ interface to enable a form of coupling between magnetism and ferroelectricity.

\begin{figure}[h]
  \includegraphics[width=8.5cm]{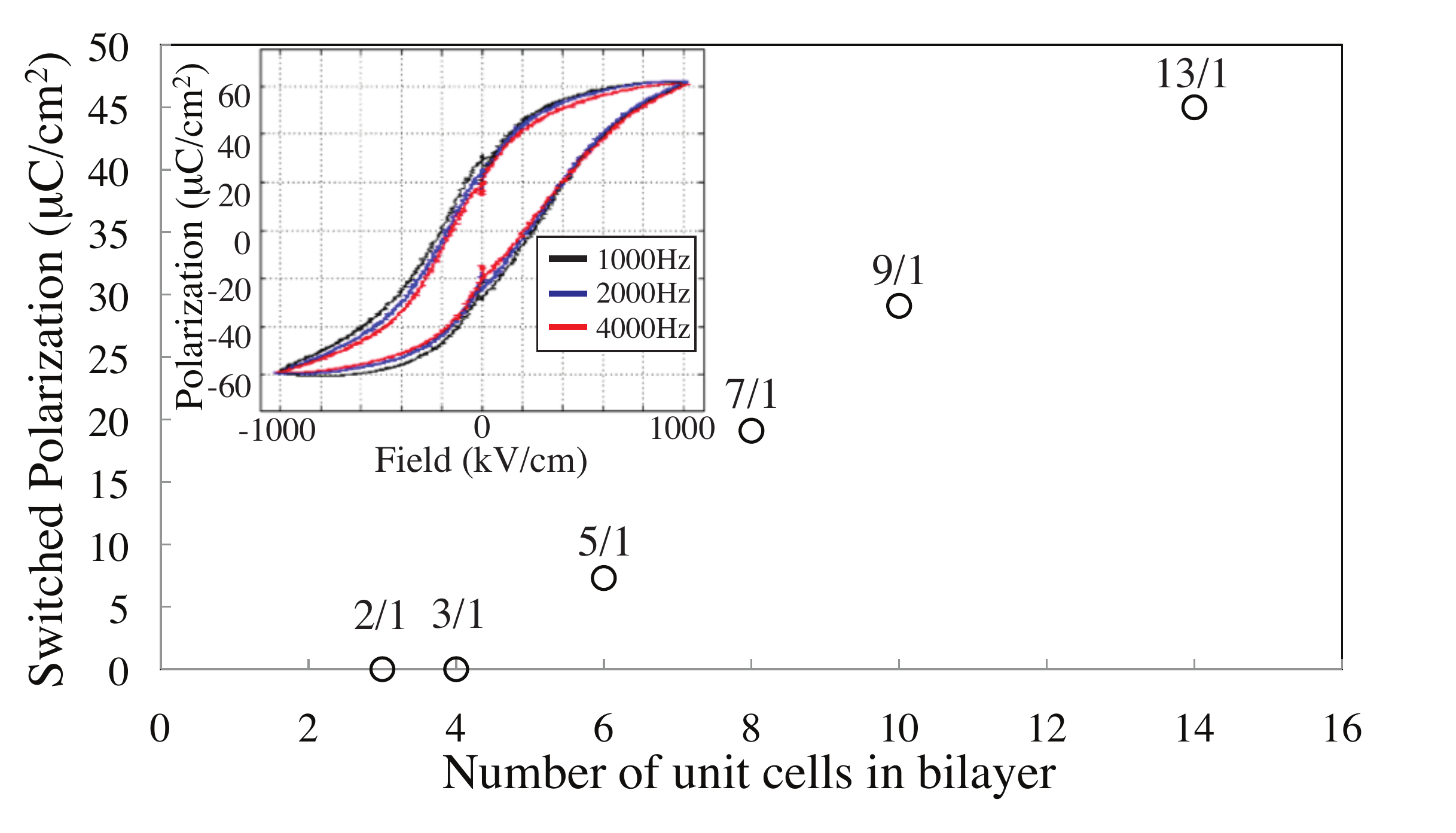}
  \caption{\textit{Switched polarization plotted as a function of the total number of unit cells in each bilayer. Inset: Polarization-field hysteresis loops measured at 3 different frequencies on a 7/1 sample.  }}\label{polfig}
\end{figure}

Experimental values for the switched ferroelectric polarization  of the samples were obtained from polarization-electric field hysteresis loops  performed on a number of samples.
Polarization hysteresis was observed in samples with a PbTiO$_{3}$ layer thickness of 5 u.c or greater.
The experimentally measured polarization as a function of the total number of unit cells in each bilayer is shown in Fig. \ref{polfig}, along with characteristic loops measured at 3 different frequencies on the 7/1 shown as an inset.
Successful hysteresis loops confirm that the thin layers of SrRuO$_{3}$ in the material are acting as dielectric layers, and allow a continuous polarization in the structure. 
An independent confirmation of ferroelectricity comes from x-ray diffraction reciprocal space maps around superlattice Bragg  peaks, shown in the supplemental information\cite{SI}.
These show diffuse scattering from the in-plane periodicity of stripe domains with polarization oriented up and down with respect to the substrate, and are similar to those seen in PbTiO$_{3}$/SrTiO$_{3}$ superlattices.\cite{Zubko10,Jo11}.
 These domains features were observed in the 7/1, 9/1, and 13/1 superlattices, but not in the 5/1 superlattice.
The lack of domain features in the 5/1 PbTiO$_{3}$/SrRuO$_{3}$ superlattice may be due to the relative instability of one polarization state with respect to each other, ie., while the polarization can be switched under field from one direction to the other, its equilibrium configuration is dominated by a single polarization direction.
Direct comparison between the calculated stable polarizations in the layer and the experimentally measured switched polarizations are difficult to make, as the two quantities, while related, are not  identical. 
However, it can be seen that the polarization values  and their dependence on the number of unit cells in each bilayer are qualitatively similar for experiment and the 3 theoretical cases considered in Table \ref{secondtable}. 
For an equivalent composition the Ru excess superlattices have higher polarizations than the Sr excess superlattices. In the ideal case marked differences in the polarization value for the up and down state become noticeable for superlattices whose bilayers contain less than 7 unit cells of PbTiO$_{3}$.
When the PbTiO$_{3}$ layer thickness is 3 layers or less the samples become fairly conductive in the out of plane direction, both in theory and experiment.

\begin{figure}[h]
  \includegraphics[width=8.5cm]{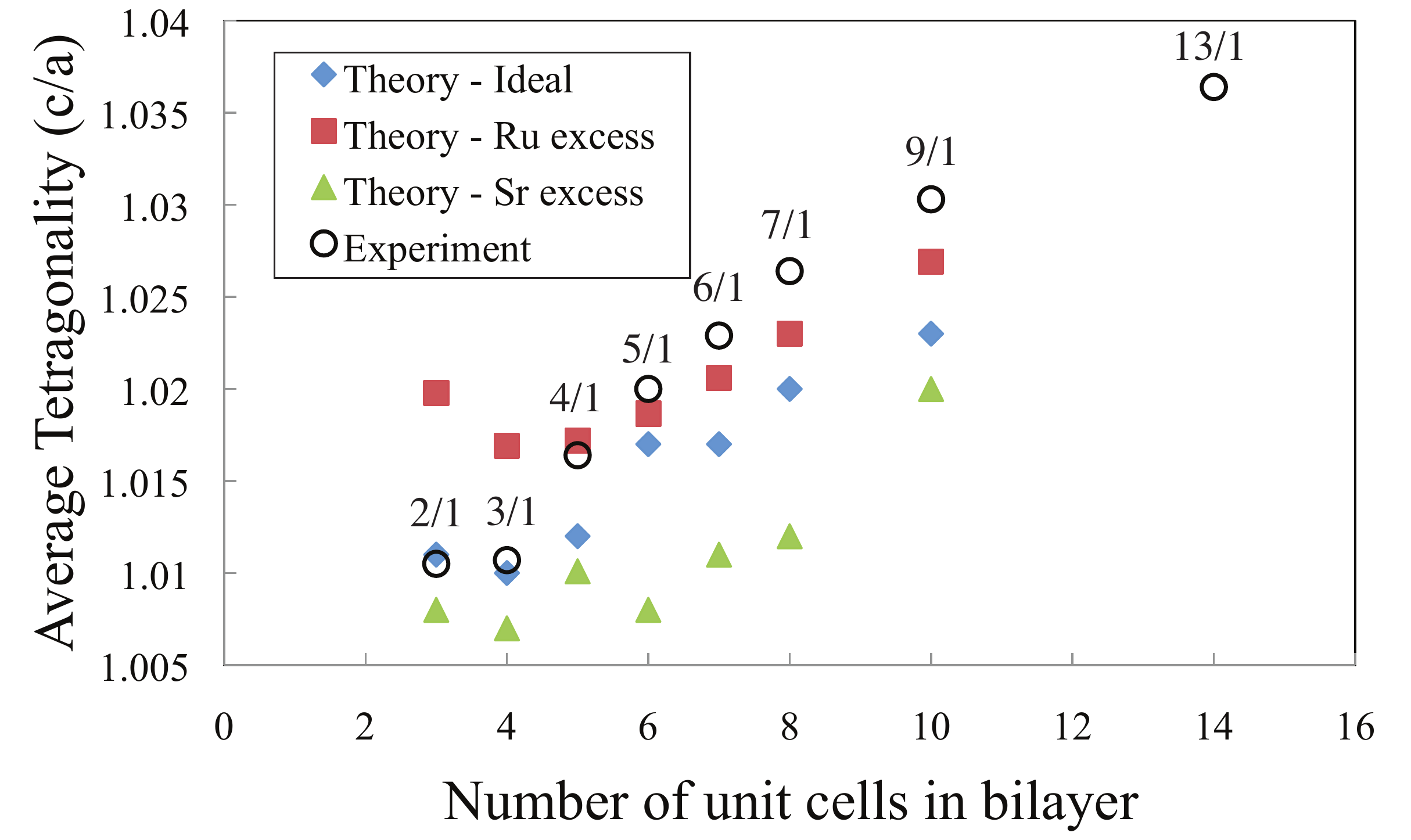}
  \caption{\textit{Average tetragonality, or  the ratio of the average out of plane lattice parameter $c$ to the lattice parameter of the SrTiO$_{3}$ substrate $a$, (i.e. $c/a$), plotted as a  function of the total number of unit cells in each bilayer.  As well as the experimental results results from DFT LDA calculations for three different kinds of interfaces are shown.}}\label{cafig}
\end{figure}

An indirect probe of the stable polarization is the average tetragonality (c/a) of ferroelectric superlattices, which can be measured using x-ray diffraction \cite{DawberPRL05,DawberAM07,Bousquet08}.
Our experimental measurements and first principles results from DFT LDA calculations are shown in Fig. \ref{cafig}. As with the polarization values in Fig. \ref{polfig}, we have plotted the results in terms of the total number of unit cells per bilayer, and included for each set of values the nominal composition. 
However, the Sr and Ru excess structures deviate from these ideal structures as described earlier; for the former one Pb-O plane has been replaced by a Sr-O plane, and for the latter a Ti-O  plane has been replaced by a Ru-O plane. 
It is difficult to make a definite conclusion which interface is present in our experimental samples solely from comparing the experimental results with the theoretical predictions shown. 
However, we suggest that a comparison of the data over the whole range of the plot would tend to exclude the Sr excess interface, and although the Ru excess interface matches the experimental data relatively well, this interface is unlikely to occur in experiment due to the high volatility of Ru.
An interesting point in this figure is that the average tetragonality begins to increase again as the number of unit cells in each bilayer is decreased.
This is because, in contrast  to PbTiO$_{3}$/SrTiO$_{3}$ where SrTiO$_{3}$ grown on SrTiO$_{3}$ has a tetragonality of 1 and the c/a montonically decreases, as the amount of PbTiO$_{3}$ decreases,  SrRuO$_{3}$ grown on SrTiO3 has a tetragonality of 1.03. 
It appears that the c axis lattice parameter of paraelectric PbTiO$_{3}$ in these superlattices is quite low, both in experiment and theory and is certainly well below the value of approximately 1.03 that is predicted by Landau theory\cite{DawberAM07}. The upturn in the c/a value occurs for the Ru excess samples first as these are, for each sample, essentially half a unit cell closer to being SrRuO$_{3}$ than the corresponding ideal superlattice.

\begin{figure}[h]
  \includegraphics[width=8.5cm]{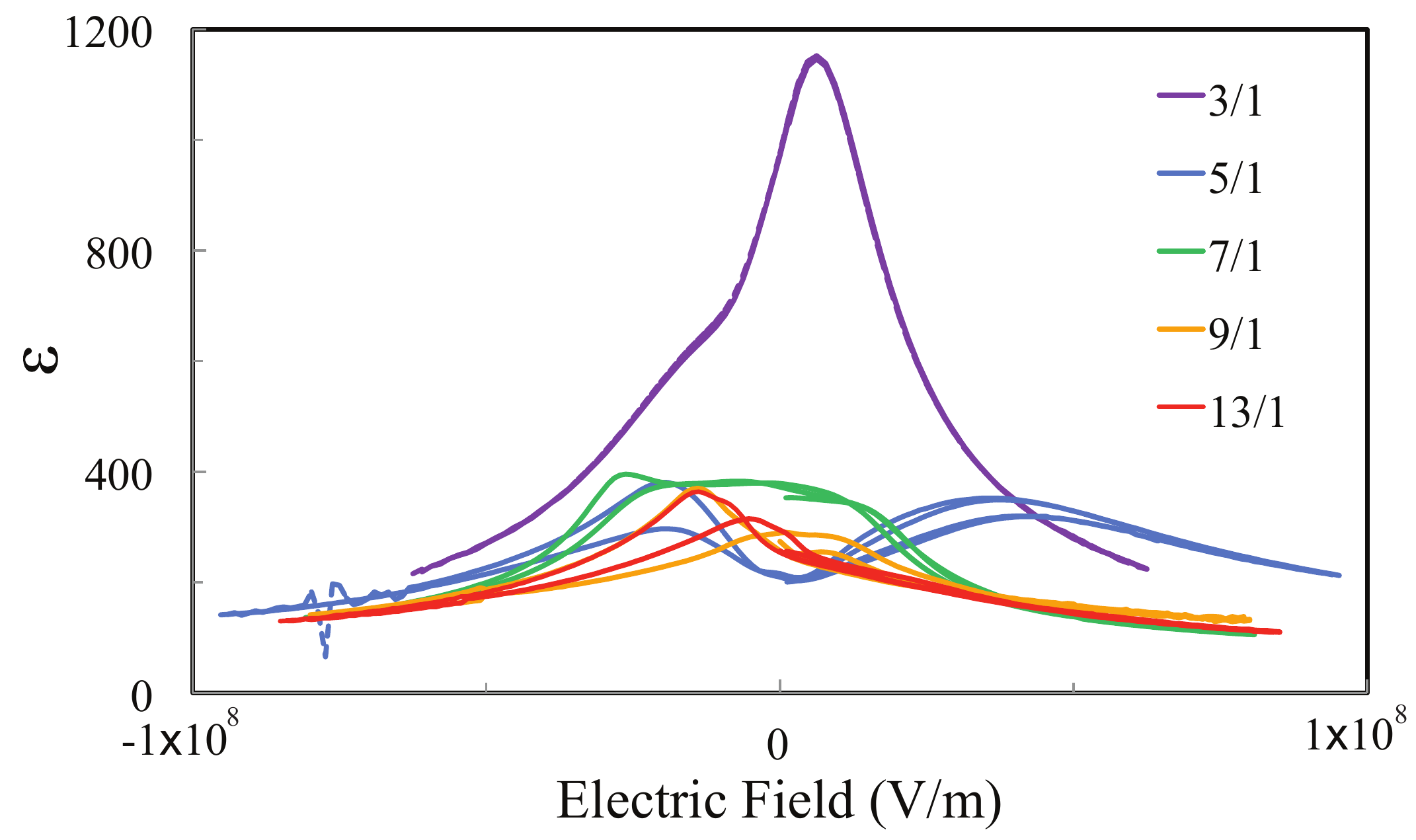}
  \caption{\textit{Dielectric constant-field loops for 5 of our samples.}}\label{ptosrodielectricexpt}
\end{figure}

The effect that the compositional breaking of inversion symmetry has on functional properties is most evident in the dielectric response of the samples.
In Fig.~\ref{ptosrodielectricexpt}, we show dielectric constant measured as a function of electric field (at a frequency of 10kHz). 
The measurements show an  evolution from a typical butterfly loop for PbTiO$_{3}$ rich samples (13/1 and 9/1 samples) to a highly asymmetric curve with greatly enhanced peak dielectric constant for the 3/1 sample. 
An unusual characteristic where two peaks are seen on each voltage trace displayed in the 5/1 sample.
The composition at which the transition from conventional ferroelectric behavior  occurs (approx. 7/1) matches the composition highlighted by theory at which compositional inversion symmetry breaking becomes a dominant factor.
To summarize our picture of this system: The 3/1 sample can be characterized as a spontaneously polarized non-switchable insulator (or, in other words, an interfacially driven pyroelectric).  Samples between 7/1 and 3/1 are best described as asymmetric ferroelectrics.  Samples with PbTiO$_{3}$ layers thicker than 7 unit cells, while still containing interfaces which compositionally break inversion symmetry are not greatly affected by them.

Besides the direct results presented here, our study offers a general demonstration of the possibilities unlocked by expanding the material set used in ferroelectric superlattices.
Our findings should motivate a broader exploration of candidate materials for the development of new artificially layered ferroelectrics.
In particular, besides the self-poling behavior that compositional broken inversion symmetry produces, the use of thin metallic materials as dielectric layers has intriguing potential for the development of highly coupled multiferroics.

\begin{acknowledgments}
We acknowledge very useful discussions with Phil Allen. MD acknowledges support from the National Science Foundation under DMR1055413.  MD and DS  acknowledge support from a SBU/BNL seed grant. MVFS acknowledges support from DOE award DEFG02-09ER16052. Use of the Center for Functional Nanomaterials, at Brookhaven National Laboratory, was supported by U.S. Department of Energy, Office of Basic Energy Sciences, under Contract No. DE-AC02-98CH10886. 
\end{acknowledgments}

\end{document}